\documentclass[12pt]{article}
\usepackage{bbm,bm}
\usepackage{axodraw,color,dsfont,cancel}
\usepackage{amsmath}
\usepackage{amssymb}
\usepackage{exscale}

\font\fourteenbf=cmbx12 scaled\magstep1

\def\a0size{6}

\newcommand{\lsi}{\raise0.3ex\hbox{$<$\kern-0.75em\raise-1.1ex\hbox{$\sim$}}}
\newcommand{\gsi}{\raise0.3ex\hbox{$>$\kern-0.75em\raise-1.1ex\hbox{$\sim$}}}

\renewcommand{\vec}[1]{{\bm #1}}

\newcommand{\be}{\begin{equation}}
\newcommand{\ee}{\end{equation}}

\begin{document}
 
\setlength{\baselineskip}{0.6cm}
\newcommand{\figysize}{16.0cm}
\newcommand{\figtopspace}{\vspace*{-1.5cm}}
\newcommand{\figbottomspace}{\vspace*{-5.0cm}}
  
% don't use the following line with revtex
%\renewcommand{\theequation}{\thesection.\arabic{equation}}
%\newcounter{saveeqn}
%\newcommand{\alphaeqn}{\refstepcounter{equation}\setcounter{saveeqn}{\value{equation}}%
%\setcounter{equation}{0}%
%\renewcommand{\theequation}{%
%        \mbox{\thesection.\arabic{saveeqn}\alph{equation}}}}%
%\newcommand{\reseteqn}{\setcounter{equation}{\value{saveeqn}}%
%\renewcommand{\theequation}{\thesection.\arabic{equation}}}

\begin{titlepage}
\begin{flushright}
BI-TP 2010/10
%\\
%\today
%\\
\end{flushright}
\begin{centering}
\vfill

{\fourteenbf 
\centerline{ 
   Hard thermal loops for  soft or collinear external momenta
      }
}

\vspace{1cm}

Denis Besak\footnote{dbesak@physik.uni-bielefeld.de} and
Dietrich B\"odeker\footnote{bodeker@physik.uni-bielefeld.de}

\vspace{.6cm} { \em 
Fakult\"at f\"ur Physik, Universit\"at Bielefeld, D-33615 Bielefeld, Germany
}

\vspace{2cm}
 
{\bf Abstract}

\end{centering}
 
\vspace{0.5cm}
\noindent
We consider finite temperature 1-loop diagrams with hard loop momenta
and an arbitrary number of external gauge fields when the external
momenta are either soft, or near the light cone and nearly collinear
with the loop momentum. We obtain a recursion relation for these
diagrams which we translate into an equation for their generating
functional.  By integrating out the soft fields while keeping two
collinear ones we find an integral equation, originally due to Arnold,
Moore, and Yaffe, which sums the bremsstrahlung and pair annihilation
contribution to the thermal photon production rate.

\vspace{0.5cm}\noindent

%PACS numbers: 11.10.Wx, 11.15.-q
%Keywords: finite temperature, real time correlation functions,
%classical limit, quantum corrections
 
\vspace{0.3cm}\noindent
 
\vfill \vfill
\noindent
 
\end{titlepage}

%%%%%%%%%%%%%%%%%%%%%%%%%%%%%%%%%%%%%%%%%%%%%%%%%%%%%%%%%%%%%%%%%
\section{Introduction}
\label{sc:introduction} 
%%%%%%%%%%%%%%%%%%%%%%%%%%%%%%%%%%%%%%%%%%%%%%%%%%%%%%%%%%%%%%%%%

Particle production in  thermal and non-thermal systems plays an
important role in relativistic heavy ion physics and in cosmology. Typically 
there is some weakly interacting particle species with a low abundance in a
hot medium, from which these particles are then produced. Examples are photons
produced in a quark--gluon plasma, or dark matter candidates such as
gravitinos, axions, or axinos produced in an otherwise thermal universe and at
reheating after inflation. 

The production of photons from a thermal quark-gluon plasma is remarkably
complicated. At {\em  leading logarithmic order}, that is, at leading order not just
in the strong coupling $ \alpha  _ { \rm S }   $, but rather in $ \alpha  _ { \rm S }
\log ( 1/\alpha  _ { \rm S } ) $ only $     2 \to 2 $ scattering processes
contribute and the rate has been computed by Kapusta {\em  et al.}
\cite{kapusta} and by Baier {\em  et al.} \cite{baier}.
However, already for the complete leading order contribution 
 the production cannot simply be understood in terms of scattering
processes involving only a handful of particles. Bremsstrahlung and pair 
annihilation  involve multiple interactions via soft gluon exchange. 
Such processes are not suppressed, despite the large number
of interactions \cite{aurenche,arnoldPhoton}. The emission occurs almost collinearly so that
internal lines are nearly on-shell and compensate the suppression. Viewed in
position space the radiated photon and its source overlap over large distances, and the 
interference of different interactions
cannot be neglected. This leads to the so-called Landau-Pomeranchuk-Migdal
(LPM) effect \cite{arnoldPhoton,landau,migdal,AurencheLPM}.
The complete leading order
photon production rate has been computed at leading order by Arnold, Moore
and Yaffe by summing all relevant diagrams. 

In this paper we present a  different approach. We consider hard
particles with momenta of order $ T $ which propagate through a 
gauge field background. The gauge field momenta are  either soft ($ k \sim g T $)
or are 
almost on-shell with virtuality $ k ^ 2 \sim g ^ 2 T ^ 2 $ and almost collinear with
the loop momentum.
Formally we integrate out the hard modes 
in this soft and collinear background, leaving us with an effective theory for
soft and collinear
modes. The effective theory is described by an equation which has a similar
structure as the non-abelian Vlasov equations \cite{blaizot93} describing Hard
Thermal Loops \cite{htl}.
Then we distinguish the soft and collinear fields and identify
them with the soft gluon field and the photon field, respectively.
In a second step we integrate out
 the soft gluons  to 
 obtain an effective theory for the
collinear gauge fields only. \footnote{This approach is similar to the one
  used in
  Ref.~\cite{effective}, where soft gauge 
fields were integrated out to obtain an
effective theory for ultrasoft ($ k \sim g ^ 2 T $) fields.} 
The effective theory takes the form of an integral equation
which has been obtained previously in  
the calculation of the photon production rate \cite{arnoldPhoton}.

This paper is organized as follows. In Sec. \ref{sc:rate} we briefly recall the
relation between thermal photon production rate and the finite temperature
polarization tensor. The main part of the paper is contained in Sec.
\ref{sc:diagrams} with the calculation of the
1-loop diagrams in a soft or collinear gauge field background. The key steps
of our calculation are the approximation (\ref{propagator})  and the partial fractioning
(\ref{fractioning}). These are used both to  compute the 2-point function
(Sec.~\ref{sc:2pointfunction}) 
and then to obtain the recursion relation for the $ n $-point functions in
terms of $ (n -1) $-point functions (Sec.~\ref{sc:npoint}). Then all $ n $-point functions are put
together in an effective action (Sec. \ref{sc:current}). In  Sec.~\ref{sc:integrate} 
we integrate out the gluon fields and in Sec.~\ref{sc:fermionic} we  
discuss the generalization from
scalar quarks to spin-1/2 quarks. We summarize and conclude in
Sec.~\ref{sc:conclusions}. Finally, in 
Appendix \ref{ap:connected} we show that the connected pieces which we
encounter in Sec.~\ref{sc:integrate} vanish. 
\vspace{3mm}% \\
%non-equilibrium situations.

\noindent{\bf  Notation and conventions:} We use the metric with
signature~$ +   - - - $. 
%%%%%%%%%%%%%%%%%%%%%%%%%%%%%%%%%%%%%%%%%%%%%%%%%%%%%%%%%%%%%%%%% 
\section{The thermal photon production rate}      
\label{sc:rate} 
%%%%%%%%%%%%%%%%%%%%%%%%%%%%%%%%%%%%%%%%%%%%%%%%%%%%%%%%%%%%%%%%% 
The rate per unit volume at which real photons with momentum $ \vec k $ and polarization vector $
\varepsilon  _ \mu  $ are produced in a hot QCD gas can be written as 
\be \label{prod_rate}
  ( 2 \pi  ) ^ 3 2 k  ^ 0 \frac{ d \Gamma  }{d ^ 3 k} = 
   e ^ 2 \varepsilon    _  \mu 
   \varepsilon    _  \nu ^ * \Pi  ^  { \mu  \nu  } _ < ( k ) 
\ee 
where $ k ^ 0 = |\vec k | $ is the photon energy and 
\begin{align} 
   \Pi  ^  { \mu  \nu  } _  < ( k ) =
 \int d ^ 4 x e ^{ i k \cdot x }  \langle  J ^  \mu  ( 0 ) J ^ \nu ( x )
   \rangle 
   \label{2point}
\end{align} 
is a 2-point function of electromagnetic current operators $ J ^ \mu  $.
We have not included the electromagnetic coupling constant in $ J^\mu $, which
therefore appears in Eq.~(\ref{prod_rate}). The 2-point function (\ref{2point}) 
can be written in terms of the retarded polarization tensor
\begin{align} 
   \Pi  ^  { \mu  \nu  }  _ < ( k ) = 2 f _ { \rm B } ( k ^ 0 ) {\rm Im } \Pi
   ^    { \mu  \nu  } _ { \rm ret } ( k ) 
   \label{retarded}
\end{align} 
where $ f _ { \rm B } ( E ) = 1/( e ^{ E/T } -1 ) $ is the Bose-Einstein
distribution at temperature $ T $. 

Among the leading order contributions to the production rate (\ref{prod_rate})
are  QCD Compton and $ q \bar q $ 
annihilation. These are $ 2 \to 2 $ scattering
processes for which only a few diagrams contribute. 

We will only be  
concerned with bremsstrahlung and pair annihilation
contributions. There a quark undergoes multiple scattering with 
 soft momentum transfer $ q \sim g T $ 
 and at some point emits a bremsstrahlung photon or annihilates with an
antiquark into a photon. 

Therefore we want to evaluate $ \Pi ^  { \mu  
  \nu  } $ with a quark interacting via soft gluons only.  For simplicity we first consider
scalar quarks, the generalization of our method to spin-1/2 quarks is
straightforward and is described in Sec.\ref{sc:fermionic}. The
first step we will take is to  integrate out the scalar 
quark fields in a soft or collinear gauge field  background. 
That means that we have to calculate diagrams with two external photon lines
and an arbitrary number of 
soft external gluon lines.

%%%%%%%%%%%%%%%%%%%%%%%%%%%%%%%%%%%%%%%%%%%%%%%%%%%%%%%%%%%%%%%%% 
\section{1-loop diagrams with soft or collinear external gauge fields}      
\label{sc:diagrams} 
%%%%%%%%%%%%%%%%%%%%%%%%%%%%%%%%%%%%%%%%%%%%%%%%%%%%%%%%%%%%%%%%%
We consider thermal 1-loop diagrams
with an arbitrary number of external 
gauge field legs. For the bremsstrahlung and pair annihilation contribution to
photon production, two of the gauge field momenta are hard ($ k \sim T $) and
correspond to the produced photon. The  remaining ones are 
soft ($ k \sim gT $) gluons. The particle in the loop 
corresponds to a quark which suffers soft scattering via gluon exchange and 
radiates the photon. 

Even though the photons
and the gluons play very different roles,  our approach allows for a
unified treatment, and in this section we do not have to distinguish these two. 
First we review the relevant kinematics which has been
extensively discussed in Ref.~\cite{arnoldPhoton}. It allows us 
to simplify propagators and
vertices. We obtain a 
recursion relation between a diagram with $ n $ external gauge field lines
and the difference of two diagrams with $ n -1 $ external lines. 
 Then we consider the current
induced by the gauge fields which is the first derivative of
the generating functional of all   $ n $-point
functions. The recursion relation turns into a generalized kinetic equation. 
It can be viewed as a generalized Vlasov equation which contains
a convective term and a force term. 

%%%%%%%%%%%%%%%%%%%%%%%%%%%%%%%%%%%%%%%%%%%%%%%%%%%%%%%%%%%%%%%%%
\subsection{Kinematics and power counting}
\label{sc:kinematics}
%%%%%%%%%%%%%%%%%%%%%%%%%%%%%%%%%%%%%%%%%%%%%%%%%%%%%%%%%%%%%%%%%
The hard momenta we are considering are all almost collinear. Up to
higher orders they all point into the same direction which we denote by $ \vec
v $ with $ \vec v ^ 2 = 1 $. The 3-momentum components in the $ \vec v
$-direction are denoted by
\begin{align} 
    p _ \| \equiv \vec p \cdot  \vec v
\end{align} 
We define the light-like vector  $ v \equiv 
( 1,  {\vec v }  ) $. One has to account for three distinct momentum scales. 
\begin{enumerate}
\item The emitting charged particle, which corresponds
to the particle in the loop, and the emitted particle both have  $ p _ \| $ 
of order $ T $, which is our hard scale.
\item All 3-momenta perpendicular to  $ \vec v
$ are soft, $ \vec p _\perp \sim g T $. Furthermore, all momentum components
 of the gluons are
soft, 
$ k _ \mu  \sim g T $.%, and the gluon virtuality is soft, $ k ^ 2\sim g T ) ^ 2 $. 
\item Finally, all 4-momenta $ k $ are `collinear', 
$ v \cdot k = ( k _ 0 - k _ \| ) \sim g ^ 2 T $.  
\end{enumerate}
This includes the case that the emitted photon is off-shell by an amount 
$ k ^ 2 \sim g ^ 2 T ^ 2 $ which is relevant for dilepton production. 

All momenta in the loop have $ k ^ 2 \sim g ^ 2 T^2 $. Therefore the propagators
are sensitive to the  so-called 
asymptotic mass $ m \sim g T $, \footnote {It is
  oftentimes referred to as $ m_\infty  $. } 
which is given by the real part of the thermal self-energy, with hard loop momentum, of a
light-like hard particle \cite{Flechsig}. It is thus generated by integrating out the gluons
with hard momenta. This procedure does not yet yield a thermal width
which is only generated by integrating out the gluons with soft
momenta. \footnote{Therefore, in contrast to \cite{arnoldPhoton} we
  never include the (IR divergent) thermal width in the quark propagators,
  only the asymptotic mass.} For scalars and fermions in the representation $ r $ of the
gauge group  
\be
    m^2 = \frac{1 }{4} C_2(r) g^2 T^2
   \label{mass}
\ee
with the quadratic Casimir of the representation $ r $. In a
SU($N$) gauge theory   $
C_2(r) = ( N ^ 2 -1 ) /( 2 N ) $. 

%%%%%%%%%%%%%%%%%%%%%%%%%%%%%%%%%%%%%%%%%%%%
\subsection{2-point function}
\label{sc:2pointfunction}
%%%%%%%%%%%%%%%%%%%%%%%%%%%%%%%%%%%%%%%%%%%%

In this section we explicitly compute the 2-point function. It turns out that 
all $n$-point functions with $n > 2$ can be obtained from $ (n -1) $-point
functions through a simple recursion relation.
Furthermore, all kinematic approximations which are needed for
the general case already appear for the 2-point function. 

We work in the imaginary time formalism, where the loop integral consists
of a sum over imaginary Matsubara frequencies and an integral over 3-momenta. 
The external momenta must also be taken imaginary and can be continued to real
values only after the sum has been performed. 
The Matsubara sum is performed as usual by writing it as a contour integral
\begin{align}
        T \sum_{ p _ 0 = i n 2 \pi  T} h( p _ 0 ) 
        =  
        \int_C \frac{d p _ 0  }{2\pi i}
        \left [ \frac12 + f _ { \rm B }  (p _ 0  )    \right ] h(p _ 0  ) 
        \label{matsubara}
\end{align} 
where the integration contour goes up on the right of the imaginary axis
and goes down on the left of the imaginary axis. Then we close 
the contour around the poles of the propagators.
After evaluating the integral \eqref{matsubara},
we can continue $ k _ 0 $ from a Matsubara frequency towards the real
axis. The retarded polarization tensor  is obtained by taking 
$ k_ 0 = \mbox{Re} (  k_ 0 ) + i \epsilon $. 

Now we describe the two main steps of our calculation. In the imaginary time
formalism  the scalar propagator appearing in  Fig.~\ref{fg:npoint} is 
\begin{align} 
  \Delta   ( p  ) \equiv \frac{ -1}{ p  ^ 2 - m  ^ 2 } 
   = \frac{ -1}{ \bar v \cdot p \,\,  v \cdot p - \vec p _\perp ^ 2 - m  ^ 2 } 
%   =  \frac{ - 1}{   
%   ( p ^ 0 + E _ { \vec p  }  )}
   \label{propagator}
\end{align} 
with $ \bar v = (1, -\vec v ) $ . 
The other propagator in 
 Fig.~\ref{fg:npoint}(a) will only be parametrically large, of order $ g ^ {-2 } 
T ^{ -2 } $, when $ v \cdot p \sim v \cdot k $
because $ v \cdot k \sim g ^ 2 T $. In this case 
$ p _ 0 $ approximately equals $ p _ \| $, and $ \bar v \cdot p \simeq 2 p 
_ \| $. Therefore we can approximate 
\begin{align} 
   \Delta  ( p ) \simeq 
    \frac{ 1}{ 2 p _ \|} D ( p ) 
   %= \frac{ 1 } { 2 p _ \| } 
   %\frac{ -1}{ v \cdot p  
   %            - \frac{  \vec p _\perp ^ 2 + m  ^ 2 }{2 p _ \| }   
   %           } 
    \label{propapp}
\end{align} 
with
\begin{align} 
   D ( p ) 
   = 
   \frac{ -1}{ v \cdot p  
               - (   \vec p _\perp ^ 2 + m  ^ 2 ) / ( 2 p _ \| ) 
              } 
    \label{D}
\end{align} 
After this approximation the propagator, viewed as a function of the energy
variable, has only one pole.  \footnote { Note that we have made the
  approximation (\ref{propapp}) {\em before} the Matsubara summation, that is,
  when $ p _ 0 $ is still purely imaginary and of order $ T $, even though
  (\ref{propapp}) is valid only for real frequencies. This is justified,
  however, because the second pole in $ p _ 0 $ of the propagator, which gets
  lost through the approximation (\ref{propapp}), would give $ v \cdot p
  \simeq -2 p _ \| \sim T $. With this value of $ v \cdot p $ the second
  propagator $ \Delta ( k - p ) $ would be of order $ T ^ {-2 } $. This is
  suppressed relative to the contribution we keep, for which we have $ \Delta
    ( k - p ) \sim g ^{ -2 } T ^ {-2 } $. The same type of argument applies to
  the poles of $ \Delta ( k - p ) $.  }  For the following it turns out to be
very convenient to partial fraction the product of the two propagators,
\begin{align} 
   D  ( p ) D ( p - k ) 
   =
   \frac1{ \epsilon   ( k, \vec p ) } 
\left [ D ( p ) 
       - D ( p - k ) \right  ]
     \label{fractioning}%51.1
\end{align} 
where
\begin{align} 
   \epsilon    ( k,  \vec p )
   \equiv 
   %[ 
     v \cdot k + \frac{  ( \vec p _\perp - \vec k _\perp  ) ^ 2 + m  ^ 2} 
     { 2 (  p _ \| - k _ \| ) }
   - \frac{ \vec p _\perp ^ 2 + m  ^ 2} 
     { 2  p _ \| } 
   %\right ] ^{ -1 } 
    \label{defoflambda}
\end{align}
denotes the difference of the poles  of the two propagators which is of order 
$   g ^ 2 T $. 

Both vertices are proportional to  $ ( 2 p - k ) _ \mu  $. We
associate the factor  $ ( 2 p _ \| ) ^{ -1 } $ of Eq.~(\ref{propapp}) with the left vertex in
Fig.~\ref{fg:npoint}(a) rather than with the propagator itself 
and define the resulting vertex factor
\begin{align} 
  V ^ \mu  ( p, p -k ) \equiv \frac{ 1 } { 2 p _ \| } ( 2 p - k ) ^  \mu 
  \label{vertexfactor}
\end{align} 
The different components
of $ V $  have different orders of magnitude. The largest component is in the
direction of $ v $ and is $ O ( 1 ) $.  
The transverse components are of order 
$ g  $, and the $ \bar v   $-component is  $ O ( g ^ 2  ) $. For 
computing the production rate of real photons one needs the transverse 
components. It is therefore not sufficient to take into account only the leading
order piece. We also include  the  transverse components, and the approximation 
\begin{align} 
    V ^ \mu  ( p, p -k )    \simeq 
   \frac{ 1 } { 2 p _ \| } \left [ 
   ( 2 p _ \| - k _ \| ) v ^  \mu  + ( 2 p - k ) _\perp ^ \mu 
   \right ] 
    \label{vertex}
\end{align} 
is implicitly understood in the following. 

We write the polarization tensor as
\begin{align} 
  \Pi  _ { \mu  \nu  } ^{ ab } ( k )   = 
  \int \frac{ d ^ 3 p } { ( 2 \pi  ) ^ 3 } 
  V _ \mu  ( p, p -k ) 
  {\rm tr} \left ( t ^ a \widehat \Pi  _ {  \nu  } ^ b ( k ,\vec  p) \right ) 
  \label{pimunu}
\end{align} 
which turns out to be convenient for 
 discussing the general $ n $-point function, and define
\be 
\mathcal{F}(p_\parallel,k_\parallel) \equiv 
  f _ {\rm B}(p_\parallel)
   - f _ {\rm B} (p_\parallel -  k_\parallel)
   \label{bose}
\ee
Our result for the 2-point function is then
\begin{align} 
   \widehat \Pi  _ {   \nu  } ^ b ( k , \vec  p)   
   =
   \frac{ 1 } { \epsilon   ( k,  \vec p ) } 
   \mathcal{F}(p_\parallel,k_\parallel)
   V _ \nu  ( p - k, p ) 
   t ^ b
   \label{54}
\end{align}  
As we already stated below Eq.~(\ref{2point}), we do not include the gauge coupling
 in $ \Pi  _ { \mu  \nu  }$. 

There is also the tadpole diagram containing the 4-point vertex which 
contributes to the 1-loop polarization tensor. It does not
have the collinear enhancement since it does not depend on the external
momentum. But its transverse components are of the same order
as those of the diagram in Fig.~\ref{fg:npoint} (a). However, 
due to the independence on 
the external momentum the tadpole diagram
has no discontinuity and  therefore does not
contribute to the production rate. Therefore we do not consider it here. 

%%%%%%%%%%%%%%%%%%%%%%%%%%%%%%%%%%%%%%%%%%%%
\subsection{Recursion relation for $n$-point diagrams}
\label{sc:npoint}
%%%%%%%%%%%%%%%%%%%%%%%%%%%%%%%%%%%%%%%%%%%%
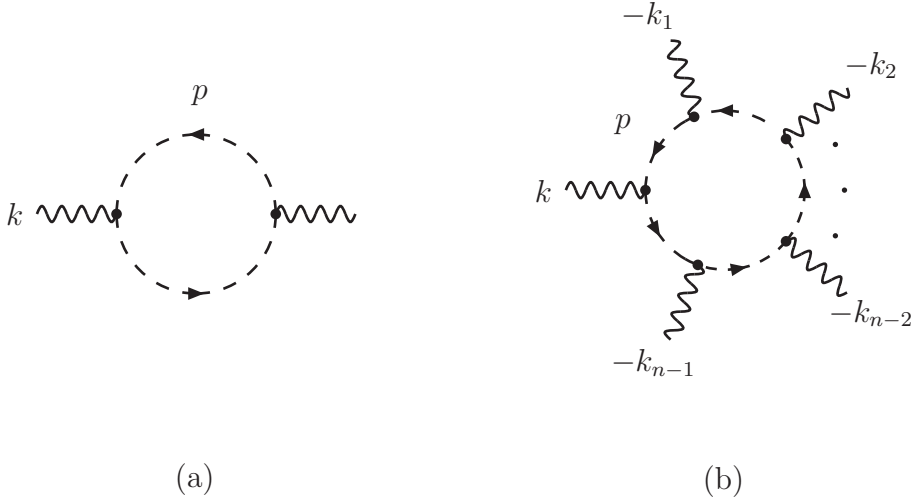
\begin{figure}[t] 
  \centering
\begin{picture}(50,90)(230,20)

\SetWidth{1}
%\SetColor{Black}

%\Photon(70,60)(120,60)(3)(6)
\Vertex(180. ,60. ){2}
\Photon(180,60)(210,60){3}{4}

\DashArrowArc(150,60)(30,0,180){5}
%\DashArrowArc(150,60)(30,90,180){5}
\DashArrowArc(150,60)(30,180,0){5}

%\Boxc(70,60)(3)(6)

\Vertex(120. ,60. ){2}
\Text(80,60)[]{ $k$}
\Photon(120. , 60.)( 90., 60.){3}{4}

\Text(150,105)[]{ $p$}

%\Text(147,15)[]{$p  - k$}
%\Text(200,95)[]{\tiny $P-Q-K$}
%\Text(110,95)[]{\tiny $P-Q$}

\Text(150, -40)[]{(a)}

\end{picture}

\begin{picture}(50,90)(30,-80)

\SetWidth{1}
%\SetColor{Black}

%\Photon(70,60)(120,60)(3)(6)
%\Photon(180,60)(230,60){3}{6}
\Text(80,60)[]{ $k$}
\Photon(120. , 60.)( 90., 60.){3}{4}
\Vertex(120. ,60. ){2}
\DashArrowArc(150,60)(30,120,180){5}
\Vertex(138.343 ,87.6427 ){2}
\Text(119.479,126.382)[]{ $-k _ 1 $}
\Photon(129.479,116.382)(139.739,88.1908){3}{4}
\DashArrowArc(150,60)(30,60,120){5}
\Vertex(172.98 , 79.28 ){2}
\Text(203.96 , 107.57 )[]{ $-k _ 2 $}
\Photon(172.98 , 79.28 )(195.96 , 98.57 ){3}{4}
\DashArrowArc(150,60)(30,320,40){5}
\Vertex( 172.98, 40.7){2}
\Text(203.96 , 13.4 )[]{ $-k _ { n -2 }  $}
\Photon( 172.98, 40.7)(195.96 , 21.4 ){3}{4}
\DashArrowArc(150,60)(30,240,320){5}
\Vertex( 139.739, 31.8092){2}
\Text(121.479 , -5.61844)[]{ $-k _ { n - 1 }  $}
\Photon(139.739 ,31.8092 )(129.479 , 3.61844){3}{4}
\DashArrowArc(150,60)(30,180,240){5}

% here come the dots:
\Vertex(191.575 , 77.2208){1}
\Vertex( 195., 60.){1}
\Vertex( 191.575, 42.7792){1}
%\Vertex( , ){1}]
%\Vertex( , ){1}
%\Vertex( , ){1}

%\Gluon(150,140)(150,90){3}{5}
%\Text(163,117)[]{$k$}

%\Vertex(120,60){5}
%\Vertex(180,60){2}
%\Vertex(150,90){2}

%\Text(60,60)[]{$\varepsilon$}
%\Text(241,61)[]{$\varepsilon^*$}
%\Text(203,72)[]{ $q$}

\Text(110,85)[]{ $p$}
%\Text(186,89)[]{ $p-k$}

%\Text(105,30)[]{$p  - k$}
%\Text(200,95)[]{\tiny $P-Q-K$}
%\Text(110,95)[]{\tiny $P-Q$}

\Text(150, -50)[]{(b)}

\end{picture}

\caption{1-loop diagrams with soft or collinear external gauge field lines. 
  Only the 2-point function (a) needs to be calculated explicitly.
  The $ n $-point functions (b) are related to the $ ( n -1 )  $-point functions 
  by a recursion relation. All external momenta are
  outgoing. 4-momentum
conservation implies
$
   k = \sum _ { j = 1 } ^ {n - 1} k _ j 
$.}
\label{fg:npoint}
\end{figure}
%%%%%%%%%%%%%%%%%%%%%%%%%%%%%%%%%%%
Now we consider the  diagram with  $n > 2$ external gauge field lines in
Fig.~\ref{fg:npoint}(b). 
We will find that it can be recursively related to 
 diagrams with $ n - 1 $ external lines. To obtain
this relation we pick out one vertex, the leftmost one in
Fig.~\ref{fg:npoint}(b) which carries momentum $ k $. 
The remaining ones have incoming momenta $ k _ j $
with $ j = 1, \ldots , n - 1$. 
As for the 2-point function we can use the approximation 
(\ref{propapp}) for the propagators. 
After the partial fractioning (\ref{fractioning}) 
this diagram is proportional to the difference of the diagrams in which either
the vertex with $ k _ 1 $ or the vertex with $ k _ { n -1 } $ is omitted.

Each vertex carries a generator $ t ^ a $ of some gauge group. As before we do
not include the gauge couplings in the vertices.  From the calculation of the
2-point function we know that in the vertices we have $ p ^ 0 \simeq p _
\| $.  In analogy with Eq.~(\ref{pimunu}) we write
\begin{align} 
   \Pi  ^ { ( n ) } { } ^{ a a _ 1 \cdots  a _ {n - 1} } _ { \mu  \mu  _ 1 \cdots
     \mu   _ {n - 1} } (  k _ 1, \ldots  , k _ { n-1 } )  
   = \int \frac{ d ^ 3 p } { ( 2 \pi  ) ^ 3 } V _ \mu  ( p, p - k ) 
   {\rm tr} \left [ t ^ a 
   \widehat \Pi  ^ { ( n ) }  
   { } ^{  a _ 1 \cdots  a _ {n - 1} } _ {  \mu  _ 1 \cdots
     \mu  _ {n - 1} } ( k _ 1, \ldots  , k _ { n-1 }, \vec p ) \right ]
   \label{52}
\end{align} 
 
We only have to
consider the two propagators $ D  ( p ) $ and $ D  ( p - k ) $ which
are connected to the left vertex in Fig.~\ref{fg:npoint}(b). 
We apply the same approximations as for the
2-point function including the
partial fractioning (\ref{fractioning}). The two terms in Eq.~(\ref{fractioning}) 
are then proportional  to diagrams in which either the propagator with
momentum $ p $ or the one with momentum $ p - k $ has been omitted from
Fig.~\ref{fg:npoint}(b). Thus, if one also leaves out the vertex factors 
connected to these propagators each of the two terms gives a
 ($ n -1 $)-point function.
For the second term in Eq.~(\ref{fractioning}) we obtain a contribution in 
which the propagator with momentum $ p $ has been
omitted. 
Here we perform a shift in the summation variable, $ p ^ 0 \to p ^ 0 
+ k _ 1 ^ 0 $. Then the remaining propagators are the same which appear in the 
$ ( n - 1 ) $-point function with incoming momenta $ k _ { 2 } , \ldots  
, k _ { n  - 1 } $, but with the loop 3-momentum $ \vec p $ replaced by 
$ \vec p - \vec k _ 1 $. 
Therefore  we can write $ \widehat \Pi  ^{ ( n ) } 
 ( k _ 1, \ldots  , k _ {n - 1} , \vec p ) $ 
in terms of the difference of 
$ \widehat\Pi  ^{ ( n - 1 ) } $ with either $ k _ 1 $ or $ k _ { n - 1 } $
omitted, 
\begin{align}
%\begin{eqnarray} 
 % \lefteqn{
          \epsilon    ( k, \vec p ) 
   \widehat \Pi  ^ { ( n ) }  { } ^{  a _ 1 \cdots  a _ {n - 1} } 
     _ {  \mu  _ 1 \cdots  \mu  _ {n - 1} } 
    & ( k _ 1, \ldots  , k _ {n - 1} , \vec p ) 
  %}
  \nonumber \\  = 
      &   - 
      \widehat \Pi  ^ { ( n - 1 ) }  { } ^{  a _ 1 \cdots  a _ { n - 2 }} 
   _ {  \mu  _ 1 \cdots \mu  _ { n - 2 } } 
    ( k _ 1, \ldots  , k _ { n - 2 } , \vec p ) 
       V  _ {         \mu  _ {n - 1} } ( p - k, p - k + k _ { n -1 }  )  t ^{ a _ {n -
      1} } 
  \nonumber \\ & 
    {} + V _ {\mu  _ 1 } ( p - k _ 1 , p ) 
    t ^{ a _ 1 }  
      \widehat \Pi  ^ { ( n - 1) }  { } ^{  a _ 2 \cdots  a _ {n - 1} } 
   _ {  \mu  _ 2 \cdots \mu  _ {n - 1} } 
   ( k _ 2, \ldots  , k _ { n - 1 } , \vec p - \vec k _ 1 )
   \label{53}
\end{align} 
%\end{eqnarray} 
%%%%%%%%%%%%%%%%%%%%%%%%%%%%%%%%%%%%%%%%%%%%%%%%%%%%%%%%%%%%%%%%%%%%%%%%
\subsection{The induced current}
\label{sc:current}
%%%%%%%%%%%%%%%%%%%%%%%%%%%%%%%%%%%%%%%%%%%%%%%%%%%%%%%%%%%%%%%%%%%%%%%%
Now we attach a gauge field $ W_ \mu \equiv t^a W_\mu ^a $ 
to each vertex in Fig.~\ref{fg:npoint}(b) except to the one with momentum $ k $,
and then sum over all  $ n $. 
The result can be interpreted as the current $ J _ \mu  $ which is induced by the
gauge field background. It can also be viewed as the first functional
derivative of the effective action  or generating functional of our diagrams.  
Working with the induced current is more convenient than with individual
diagrams  because one does not have to worry about summing over all
permutations of external lines. Furthermore, it can be used for integrating
out the soft gauge fields. As in Eqs.~(\ref{pimunu}) and (\ref{52}) we write
\be 
  J _ \mu  ^ a ( k ) = \int \frac{ d ^ 3 p } { ( 2 \pi  ) ^ 3 } 
  V _ \mu  ( p, p - k ) {\rm tr} \left ( t ^ a  \widehat J ( k, \vec p ) \right
  ) 
\label{defofhatJ}
\ee 
The ``unintegrated'' current $ \widehat J $ is given by 
\begin{align}  
   \widehat J ( k, \vec p ) = \sum _ { n = 2 } ^{ {\infty  } } &
   \prod _ { i = 1 } ^ {n - 1}
   \left ( \int \frac{ d ^ 4 k _ i } { ( 2 \pi  ) ^ 4 }  
W ^{ \mu  _ i } _ { a _ i } ( k _ i )  
    \right ) 
    ( 2 \pi  ) ^ 4 \delta  \left ( k - \sum _ { j = 1 } ^ {n - 1} k _ j \right ) 
    \nonumber \\ 
   \times & \widehat \Pi  ^{ ( n ) }
 { } ^{  a _ 1 \cdots  a _ {n - 1} } _ {  \mu  _ 1 \cdots
     \mu  _ {n - 1} } ( k _ 1, \ldots  , k _ {n - 1} , \vec p ) 
\end{align} 
Using Eqs.~\eqref{54} and \eqref{53} one obtains a relation 
\footnote{
Fourier transformed with respect to $ k $ 
this relation has a similar structure as the non-abelian Vlasov \cite{blaizot93} equations from
which one obtains the generating functional of Hard Thermal Loops \cite{htl}.
The linear part of the convective term $v \cdot k$ 
in the Vlasov equation got replaced by 
$\epsilon  (k,\vec p)$. 
The force term corresponds to the first term  on the RHS. Finally,
the non-linear part of the covariant convective derivative
is replaced by the integral
on the RHS.  In fact, if  one could neglect $ k $ and $ q $ relative to
$ \vec p $, then  the integral would be proportional to the Fourier transform
of the commutator  $[  v \cdot W ,  \widehat J ]$. Unlike the Vlasov equation
our equation is non-local with two sources of non-locality. One is 
the terms with $ k _ \| $ in the denominator, 
and the other is due to the fact that the second $ \widehat J $ in the 
integral depends on $ \vec p - \vec q $. 
}
for $ \widehat J ( k, \vec p )$, 
\begin{align}
  \epsilon    ( k, \vec p ) \widehat J ( k, \vec p ) 
  =   &  
   \mathcal{F}(p_\parallel,k_\parallel) 
   V (p - k, p  ) \cdot W ( k ) 
   \nonumber \\ &
    {} - \int \frac{ d ^ 4 q } { ( 2 \pi  ) ^ 4 }  
   \left [  \widehat J ( k - q , \vec p ) 
   V( p - k, p - k + q ) \cdot W ( q ) 
   \right .
   \nonumber\\ &
   \left.  
    \qquad \qquad \qquad 
   {} - V ( p - q, p) \cdot W ( q ) 
   \widehat J ( k - q , \vec p - \vec q ) 
     \right ] 
     \label{57}
\end{align} 

%%%%%%%%%%%%%%%%%%%%%%%%%%%%%%%%%%%%%%%%%%%%%%%%%%%%%%%%%%%%%%%%% 
\section{Integrating out soft gluons}
\label{sc:integrate} 
%%%%%%%%%%%%%%%%%%%%%%%%%%%%%%%%%%%%%%%%%%%%%%%%%%%%%%%%%%%%%%%%%
The photon polarization tensor which enters the production rate
\eqref{prod_rate} can be obtained from the diagrams in Fig.~\ref{fg:npoint} by 
identifying two external lines with  photons and the remaining ones with gluons.
Connecting the gluon
vertices with propagators and integrating over the gluon momenta
 will generate precisely the ladder
diagrams studied in \cite{arnoldPhoton}.  In addition, it generates quark
self-energy insertions with soft gluon loops, which at LO are purely imaginary
and correspond to a thermal width. These two contributions by themselves would
be IR divergent, but their sum is IR convergent.

In terms of the current (\ref{defofhatJ}) it means that one of the background
gauge fields is the photon field and all others are gluons. The
gluon fields are integrated out, leaving only the photon. 
Then the current is just $ \Pi ^{ \mu \nu }
A _ \nu $, from which one can read off the polarization tensor $ \Pi ^{ \mu 
  \nu } $. 

We therefore distinguish between external photon and gluon fields $ A ^{ \mu
} $ and $ G ^ \mu  $, and write
\begin{eqnarray} 
  W ^ \mu   = A ^ \mu  + G ^ \mu  
\end{eqnarray} 
The photon carries $ k _ \| $ of order $ T $, while the gluon field
has $ q _ \| \sim g T $.  In order to compute the photon
polarization tensor we consider one external photon field. In $ \widehat J $ we
only need to consider terms zeroth and first order in $ A ^ \mu  $,  $
\widehat J = \widehat J _ 0 + \widehat J _ 1 $, and what we need to compute is
$\widehat J_1$.

In the equation for $ \widehat J _ 0 $ the function $ \mathcal {F}$ 
vanishes at leading order.  Therefore $ \widehat J _ 0 $ is
suppressed compared to $ \widehat J _ 1 $ and it can be neglected in the
equation for $ \widehat J _ 1 $. Thus the equation for $ \widehat J _ 1$ takes
exactly the same form as Eq.~(\ref{57}) for $ \widehat J $, with $ W $
replaced by $ A $ in the inhomogeneous term, and with the two $ W $'s inside the
integral replaced by $ G $. We keep only the leading order piece of
the gluon vertex factors, so that $  V( p - k, p - k + q )
\simeq V (p - q, p ) \simeq v $ and thus
\begin{align}
  \epsilon    (k,\vec p) & \widehat J_1 ( k, \vec p ) 
   =   
   \mathcal{F}(p_\parallel,k_\parallel) 
   V( p - k, p ) \cdot A ( k ) 
    \nonumber \\ &
     -  \int \frac{ d ^ 4 q } { ( 2 \pi  ) ^ 4 }  
            \left [  \widehat J_1 ( k - q , \vec p ) v \cdot G ( q )  
     -  v \cdot G ( q ) 
    \widehat J_1 ( k - q , p_\parallel,\vec p_\perp - \vec q _\perp  )  \right ]  
      \label{action63}
\end{align}  

Now we would like to integrate out the gluon field. We denote the 
resulting current by $  \langle \widehat J _ 1 \rangle $. 
 In order
to see how it works we write Eq.~\eqref{action63} schematically as 
 $  \widehat J _ 1\sim A +   G \widehat J _ 1$,
leaving out all terms and 
all factors which are not relevant 
for the present discussion. Now we iterate it once to obtain 
$ \widehat J _ 1\sim A +  G ( A + G\widehat J _ 1 ) $. 
The term  $ G A $ 
vanishes when we integrate out the gluons. 
Thus we can drop this term and 
write $ \widehat J _ 1 \sim A + G G \widehat J _ 1 $.  
 Integrating out the gluons then gives 
$ \langle \widehat J _ 1 \rangle 
\sim A + \langle GG \widehat J _ 1 \rangle $. The two gluon fields can either be
contracted with each other, or with the other gluon fields in $ \widehat J _ 1 $, 
that is, $ \langle GG \widehat J _ 1 \rangle \sim \langle GG \rangle 
\langle \widehat J _ 1 \rangle + \langle GG \widehat J _ 1 \rangle _ { \rm 
connnected } $. In Appendix \ref{ap:connected} we show that the connected
part vanishes at leading order and can therefore be dropped. 
Thus, by integrating out the gluons one obtains a closed 
equation for  $ \langle \widehat J _ 1 \rangle $ of the form
$ \langle \widehat J _ 1 \rangle \sim A + 
\langle GG \rangle
\langle \widehat J _ 1 \rangle $.

Now we can become more explicit. 
After iterating the integral equation once and
integrating out the soft gluons, they have disappeared as external particles
and appear only in terms of their propagator, 
\be
 \left\langle G_\mu ^a(q) G_\nu ^b(q') \right\rangle = g ^ 2 \delta^{ab}
  \Delta_{\mu \nu}(q) ( 2 \pi  ) ^ 4  \delta(q + q') 
\label{gluonprop}
\ee
Since we are interested
in the electromagnetic current we put $ t ^ a = \mathds{1}$ in
Eq.~(\ref{defofhatJ}).  Using ${\rm tr} 
\mathds{1} = d(r)$ and $t^a t^a = C_2(r) \mathds{1}$, we obtain
\begin{eqnarray} 
\lefteqn{
  \epsilon     ( k, \vec p ) \,\, {\rm tr} \left\langle  \widehat J_1 (k,\vec p)
\right\rangle 
   = 
   d(r)    \mathcal{F}(p_\parallel,k_\parallel) 
   V (  p - k, p ) \cdot A ( k )  } \nonumber \\
  &&  {} -  2  C_2(r) g ^ 2 
  %\int  \!\!\!\!\!\!\!\!\sum _ k
   \int\frac{ d ^ 4 q } { ( 2 \pi  ) ^ 4 } 
   \frac{v^\mu
    v^\nu \Delta_{\mu \nu}(q)}{v\cdot(k - q)} {\rm tr} \left[ \left\langle
    \widehat J_1 (k,\vec p) \right\rangle - \left\langle \widehat J_1 (k,p_\parallel,\vec
    p_\perp - \vec q _\perp ) \right\rangle \right]  \hfill 
   \label{action72.1}
\end{eqnarray}
The square bracket neither depends on $ q _ 0 $ nor on $ q _ \| $, and the
integrals over   $ q _ 0 $ and $ q _ \| $
can be performed. 
 Eq.~(\ref{action72.1}) is
thus an integral equation which determines the transverse momentum
dependence $  \mbox{tr} \langle 
\widehat J _ 1 \rangle $. 
Inside the integral  we have approximated  $ \epsilon(k - q,
\vec p ')$ \\ $ \simeq v  \cdot (  k - q  ) $, i.e., we have neglected the terms containing transverse
momenta  and thermal masses, even though they are of the same order as the
term we kept. This is possible because the gluon
propagator separately depends on $ q _ 0 $ and $ q _ \| $, which are both of order $ g
T $, and not on their difference
$ v \cdot q  $, which is of order $ g ^ 2 T
$.  Therefore the terms we have omitted only contribute to a higher order
shift of the integration variable $ q _ \| $.

To 
perform the integrals over $ q _ 0 $ and $ q _ \| $ in
Eq.~(\ref{action72.1})  we use again the imaginary time formalism. It means
that we replace
the integral $( 2 \pi  ) ^{ -1 }  \int d q _ 0 $ by a sum over Matsubara
frequencies like  in  
Eq.~(\ref{matsubara}). Furthermore, it means that  $ k _ 0 $ in Eq.~(\ref{action72.1}) has to be
an (imaginary) Matsubara frequency. Only after performing the sum over $ q _ 0
$ one can
analytically continue $ k _ 0 $ towards the real axis. We
are interested in the production rate which is proportional to the imaginary
part of the retarded propagator. Therefore we have to give   $ k _ 0 $ 
a small imaginary part, i.e., $ k _ 0 = \mbox{Re} ( k _ 0 ) + i \varepsilon  $.
In Appendix
\ref{ap:connected} we show how we perform the Matsubara sum.
Using  standard results for the HTL resummed propagators
\cite{LeBellac} and the sum rule of Ref.~\cite{Aurenche1} one then obtains
\begin{eqnarray} 
   %\int \frac{ d q _ 0 \, d q _ \| } { ( 2 \pi  ) ^ 2 } 
   T\sum _ { q _ 0 } \int \frac{  d q _ \| } {  2 \pi   } 
    \,\,
   \frac{v^\mu
    v^\nu \Delta_{\mu \nu}(q)}{v\cdot(k - q)}
   \simeq 
   \frac{ i } { 4 } T \left [ - \frac{ 1 } { \vec q _\perp ^ 2} 
      + \frac{ 1 } { \vec q _\perp ^ 2 + m^2_{\rm D} } \right ] 
   \label{sumrule}
\end{eqnarray} 
where we were able to neglect the dependence on $ v \cdot k \sim g ^ 2 T
$. That is, the only dependence on $ v \cdot k $ enters through the imaginary part
of $ k _ 0 $. The Debye mass squared enters through the gluon propagators and
is given by 
\be m^2_{\rm D} = \frac{g^2 T^2}{6} (2N + N_s
+ N_f) 
\ee 
for a SU$(N)$ gauge theory
with $N_s$ complex scalars and $N_f$ Dirac fermions.

For the production rate \eqref{prod_rate} we need the polarization tensor
which we write as 
\be \label{pol_tensor}
\Pi_{\mu \nu}(k) = \int \frac{d^3 p}{(2\pi)^3} V_\mu(p , p - k) \widehat \Pi_{\nu}(k,\vec p)
\ee
The reduced polarization tensor $\widehat \Pi_\nu$ is related to $ \widehat J
_ 1  $ through
\begin{eqnarray} 
   \mbox{tr} \langle \widehat J _ 1 ( k, \vec p ) \rangle = \widehat \Pi  _ \nu  ( k, \vec p ) A
    ^ \nu  ( k )
\end{eqnarray} 
Therefore it satisfies the integral equation
\be 
\begin{aligned}
  \epsilon    (k,\vec p) & \widehat \Pi_{\nu}  (k,\vec p)   =
  d ( r ) \mathcal{F}(p_\parallel,k_\parallel) V_\nu  ( p - k, p ) 
    \hfill \\
 & +  
 i C_2(r) g ^ 2 T
   \int \frac{d^2 q _\perp }{(2\pi)^2}  
   \left [ \frac{ 1 } { \vec q _\perp ^ 2 } 
   - \frac{ 1 } { \vec q _\perp ^ 2 + m^2_{\rm D} } \right ]  
   \left[ \widehat \Pi_{\nu}(k,\vec p) - \widehat
     \Pi_{\nu}(k,p_\parallel,\vec p_\perp - 
 \vec q _\perp ) \right ]  \vphantom{\int} \hfill 
\end{aligned} 
\label{inteqphoton}
\ee
We finally want to show that our integral equation \eqref{inteqphoton} can be reduced to the one
in Ref.~\cite{arnoldPhoton} for the production of real photons. 
We choose $ \vec v $ in the direction
of $ \vec k $ so that $ \vec k _\perp = 0 $. Then we have $ v \cdot k = 0 $,
and thus $ \epsilon  ( k , \vec p ) = (  \vec p _\perp ^ 2 + m ^ 2 )k _ \|
/[2 p _ \| ( p _ \| - k _ \| ) ] $. Only the transverse
components of $\widehat \Pi_\nu$ contribute since the polarization vectors in
\eqref{prod_rate} are purely transverse. 
If we define a new function  $\vec f$ via 
\be \label{Identification}
   \widehat { \vec \Pi } _\perp(k,\vec p) = \frac{ 
   id(r)\mathcal{F}(p_\parallel,k_\parallel) }  
   {2 ( p _ \| - k _ \| ) } 
   \vec f (k,\vec p)
\ee
we obtain the equation of the same form as in \cite{arnoldPhoton} 
\begin{align}
 2 \vec p _\perp  & =  i\epsilon(k,\vec p) \vec f(k,\vec p)
  \hfill \nonumber \\ 
&  {} + g ^2 C_2(r) T
    \int \frac{d^2 q_\perp}{(2\pi)^2} \left [  \frac{1}{\vec q_\perp ^2} -
   \frac{1}{\vec q_\perp ^2 + m_ { \rm D } ^2} \right ] 
   [ \vec f(k,\vec p) - \vec f(k,p_\parallel,\vec p_\perp -
\vec q_\perp) ] \hfill 
\end{align}

%-----------------------------------------------------------------------
\section{Spin-1/2 quarks}   
\label{sc:fermionic}
%-----------------------------------------------------------------------
So far we have always been dealing with scalar quarks to avoid technical
complications. There is no need to
redo the entire calculation for spin-1/2 quarks. The results above are still valid up to a few
modifications which we now describe.

We now have to deal with
 the resummed fermion propagator 
\be
   S(p) = \frac{-1}{\cancel{p} -  \Sigma(p)}
\ee
in the high temperature limit when the zero temperature mass can be
neglected. 
In the plasma rest frame the self-energy $ \Sigma(p)$ takes the general form
\cite{weldon} 
\be \label{Sigma_Ansatz}
 \Sigma(p) =  a(p) \cancel{p} +  b(p) \gamma^0
\ee
Therefore chiral symmetry is not broken by thermal effects, and the
left- and right-chiral fermions propagate independently.  
Since $ \Sigma(p) \sim g^2 T$, we have $ a(p) \sim g^2$. 
Thus $ a ( p ) \cancel{p} $ is small compared to the tree level
contribution and may be neglected. 
For $ p \sim T $, $ p ^ 2 \sim g ^ 2 T ^ 2 $ the propagator can then be written as
\be \label{FermionProp}
   S(p) \simeq  - \frac{  \cancel{p} +  b(p) \gamma^0
                      }
                     { p^2 - 2 b(p)  p^0}
         \simeq - \frac{  \cancel{p} 
                      }
                     { p^2 - m ^ 2 }
\ee
Here  we have neglected terms of order $ g ^ 2 T $ in the numerator and 
we have identified 
\be
 b(p) = \frac{m^2}{2p^0}
\ee
at $ p ^ 2 = m ^ 2 $, where $ m $ is the asymptotic thermal mass in
Eq.~(\ref{mass}).
Since the left and right-handed fermions propagate independently, one
may consider the photon production from left-handed quarks only. The
complete rate is then twice as large. Thus one can deal with Weyl
instead of Dirac spinors, the vertices contain $ \bar \sigma ^ \mu 
$ instead of $ \gamma ^ \mu $, and the propagator (\ref{FermionProp})
contains $ \sigma \cdot p $ instead of $ \cancel{ p } $.  Up to terms
of order $ g ^ 2 T $, which we neglect in the numerator, $ p ^ \mu $ is
light-like. Therefore we can write
\begin{align} 
   \sigma  \cdot p \simeq 2 p _ \| \eta  ( p ) \eta  ^\dagger ( p )
\end{align} 
where $ \eta  ( p ) $ is a normalized eigenvector of $ \vec p \cdot \vec \sigma 
/p _ \| $ with negative eigenvalue. 
Thus we find, similarly to Ref.~\cite{arnoldPhoton},
\begin{align} 
  S _ L ( p ) 
   \simeq 
     \eta  ( p ) \eta  ^\dagger ( p ) D ( p ) 
%   { v \cdot p - ( \vec p _\perp ^ 2 + m  ^
%     2 )/( 2 p _ \| ) } 
   \label{fermionpropagator}
\end{align} 
with the same $ D ( p ) $ as in Eq.~(\ref{D}).
%%%%%
Note that, unlike in
Ref.~\cite{arnoldPhoton}, our result is valid for
both signs of $ p _ \| $. 
We associate the spinors $ \eta  ( p ) $ and $
\eta  ^\dagger ( p ) $ with the vertices on either side of the propagator
rather than with the propagator itself. Therefore the vertex factor now reads
\begin{align} 
   V ^ \mu  ( p, p - k ) = \eta  ^\dagger ( p - k ) \bar \sigma  ^ \mu  \eta
   ( p ) 
\end{align} 
instead of Eq.~(\ref{vertexfactor}). For real photon production one needs $ V
$ up to order $ g $,  
\begin{align} 
   V ^ \mu  ( p, p - k ) = v ^ \mu  + V ^ \mu  _\perp + O ( g ^ 2 ) 
\end{align} 
In the helicity basis the transverse components $ V ^ \mu  _\perp $  are
particularly simple. We choose $ \vec v $ as the 3-direction. Then 
for $ V ^{ \pm } \equiv  ( V ^ 1 \pm i V ^ 2 )/\sqrt{ 2 }  $ one
finds
\begin{align}
   V ^ + = \frac{ p ^ + } { p _ \| - k _ \| }  + O ( g ^ 2 ) , \qquad  
      V ^ - = \frac{ p ^ - } { p _ \|  }  + O ( g ^ 2 )
\end{align} 
The other difference compared to scalar quarks is due to the Fermi-Dirac
statistics, the Bose-Einstein functions get
replaced by Fermi-Dirac distributions, so that  $ \mathcal{F} $ becomes
\begin{align} 
   \mathcal{F}(p _ \|, k _ \| ) = f _ {\rm F}(p_\parallel)
   - f _ {\rm F} (p_\parallel -  k_\parallel)
   \label{fermi}
\end{align} 

%%%%%%%%%%%%%%%%%%%%%%%%%%%%%%%%%%%%%%%%%%%%%%%%%%%%%%%%%%%%%%%%%
\section{Summary and Conclusions}
\label{sc:conclusions}
%%%%%%%%%%%%%%%%%%%%%%%%%%%%%%%%%%%%%%%%%%%%%%%%%%%%%%%%%%%%%%%%%

In this paper we have obtained an integral  equation (Eq.~(\ref{57})) which sums all thermal
1-loop diagrams with an arbitrary number of soft or collinear external gauge
fields. We have applied it to compute the rate for real photon production by
bremsstrahlung and pair annihilation in a hot QCD plasma.  

Compared to the original calculation of the photon production rate in 
Ref.~\cite{arnoldPhoton} our approach is
significantly simplified by the fact the calculation is done in two
steps. In the first step we have integrated out the hard momentum modes at one
loop. The resulting effective theory is summarized by Eq.~(\ref{57}). It has a
similar structure as the non-abelian Vlasov equation which describes the Hard
Thermal Loops for soft external gauge fields. In a second step we have
integrated out the soft gauge fields corresponding to gluons. This results in the
integral equation obtained earlier in Ref.~\cite{arnoldPhoton} which sums all leading order
ladder and self energy contributions to the photon polarization tensor for
hard on-shell photons, and which thus describes the Landau-Pomeranchuk-Migdal
effect on thermal photon production.

Our approach should easily allow for generalizations. The
method can be adopted to the production of other particles than photons,
e.g. the production of spin-1/2-fermions. 
 We also hope for a possible generalization to non-equilibrium
situations, as they occur e.g. in heavy ion collision. 

%\vspace{.5cm}
{\bf Acknowledgments} 
This work was supported in part through the
DFG funded Graduate School GRK 881.

%\newpage

\appendix 
\renewcommand{\theequation}{\thesection.\arabic{equation}}

%-----------------------------------------------------------------------
\setcounter{section}0
%\textcolor{blue}{
\section{Connected pieces}   
\label{ap:connected}
\setcounter{equation}0
%-----------------------------------------------------------------------
Here we show that only the disconnected parts contribute in the calculation of
the photon production rate at leading order, as claimed in
Sec.~\ref{sc:integrate}. To simplify the discussion we can drop the dependence
on spatial momenta and on $ m ^ 2  $, and replace $ \epsilon  ( k, \vec p ) $
by $ v \cdot k $. Furthermore we leave out the vector indices of $ A $ and $ G
$. All these simplifications do not affect our argument. 

Therefore, for
the present discussion, we may study, instead of the full integral equation
\eqref{action63}, the simplified version 
\be
v\cdot k \widehat J_1(k) =    A(k) + \int_{k_1} 
G(k_1) \widehat J_1 (k - k_1) 
   \label{simple}
\ee
where we have introduced the compact notation
\be
\int_{k} \equiv T\sum_{k_0}\int \frac{d^3 k}{(2\pi)^3} 
%\quad C^\mu (k, \vec p) \equiv
%\frac{f _ {\rm B}(p_\parallel) - f _ {\rm B}(p_\parallel - k_\parallel)}{4p_\parallel
%  (p_\parallel - k_\parallel)} (2p - k)^\mu  
\ee
One can easily write down the solution  to Eq.~(\ref{simple}) 
\begin{align} 
   \widehat J_1(k) = \frac{ 1 } { v \cdot k } 
    \sum _ { N = 0 } ^ {\infty  } 
     \prod_{n = 1} ^N \left (   \int_{k_n} G ( k _ n ) 
   \frac{1}{v\cdot \left( k - \sum_ { l = 1}
    ^n k_l \right)} \right ) 
   A \left( k - \sum_{l = 1} ^N k_l \right ) 
\end{align} 
In the connected part $ \langle GG \widehat J _ 1 \rangle _ { \rm 
connnected } $, which was dropped in Sec.~\ref{sc:integrate},
the gluon field $ G ( k _ 1 ) $ is contracted with some $ G ( k _ M ) $ with $M \ge 3 $. 
Thus $ \langle GG \widehat J _ 1 \rangle _ { \rm 
connnected } $ contains  the Matsubara sum  
\be
    T\sum_{k_1 ^0} \Delta    (k_1) \frac{ 1}{v\cdot (k -
    k_1)}  \frac{1}{v\cdot (k - k_1 - k_2)} \cdots \frac{1}{v\cdot (k - k_1 
    \cdots - k_{M - 1})} 
   \label{sum}
\ee
% As in section
%\ref{sc:integrate}, we replace $ k _ 0 = \mbox{Re} ( k _ 0 ) + i \varepsilon  $ and rewrite the
%thermal sum, ignoring overall minus signs, as 
%\be 
%\left( T\sum_{k_1 ^0} \frac{v^\mu v^\nu \Delta_{\mu \nu}(k_1)}{(k_1 ^ 0 -
%    k_{1,\parallel} -i\varepsilon)^{m - 1}} \right) \cdot \ldots 
%\ee
%where we used that $k_{1,\parallel} \sim gT$ but $v \cdot k_i \sim g^2 T \ll
%k_{1,\parallel}$.
We use the spectral representation of the propagators
\be 
   \Delta(k _ 1) = - \int \frac{d\omega}{2\pi i} \frac{1}{k _ 1 ^0 - \omega}
   \operatorname{Disc} 
   \Delta(\omega, \vec k _ 1)
\ee
The thermal sum can now easily be performed using \eqref{matsubara}. Out of
the $M$ poles, only the one at $k_1 ^ 0 = \omega$ which gives $f_{\rm B}(\omega) \simeq T/\omega = \mathcal{O}( 1/g)$ 
contributes at leading order because the
gluons are soft. At all other poles the
Bose distribution function would be  $\mathcal{O}(1)$ and the corresponding contributions can
be neglected. 

After performing all Matsubara sums, we may analytically continue $ k ^ 0 $ 
towards the real axis and replace it by $ k ^ 0 + i \varepsilon  $ where 
$ k ^ 0 $ is now real.  Then (\ref{sum}) turns into
\begin{align} 
     \int \frac{d k _ 1 ^ 0 }{2\pi i}\, \frac{ T } { k _ 1 ^ 0 }
     \operatorname{Disc} \Delta  ( k _  1 ) &
   \, \frac{ 1}{ v\cdot (k -
    k_1) + i \varepsilon } \, \frac{1}{ v\cdot (k - k_1 - k_2) + i \varepsilon } 
   \nonumber \\
   &\cdots \frac{1}{ v\cdot (k - k_1 
    \cdots - k_{M - 1}) + i \varepsilon } 
\end{align} 
Here we have to integrate over the region where $ v \cdot k _ 1 $ is of order $ g ^
2 T $. That means that $ k _ 1 ^ 0 $ is equal to $ k _ { 1 \| } $ up to terms
of order $ g ^ 2 T $ and that we may replace                                    
\begin{align} 
   \frac{ 1 } { k _ 1 ^ 0 }
     \operatorname{Disc} \Delta  ( k _ 1 ) \to
   \frac{ 1 } { k _ { 1 \| }  }
     \operatorname{Disc} \Delta  (  k _ { 1\| } , \vec k _  1 )
\end{align}  
without changing the leading order result. 
Now we are done because all poles in the integrand lie above the real
$ k _ 1 ^ 0 $-axis. 
% \footnote{There is no pole 
%   at $\omega = 0$ since $\operatorname{Disc}\Delta_{\mu \nu}(\omega, \vec k_1) = 0$ for $\omega
%   = 0$.} 
Therefore we can close the integration contour  at
$ -i  \infty$ and we obtain zero.
This proves that at leading order only the
disconnected part contributes. 
 
We finally remark that if we contract $ G ( k _ 1 ) $ with $ G ( k _ 2
) $, which corresponds to the disconnected contribution (and to $M =
2$ in the calculation above), this argument fails since the integrand
does not fall off rapidly enough at infinity and the integral over a
closed loop would give a non-vanishing contribution. In fact, in that
case we obtain the result \eqref{sumrule}.

In Ref.~\cite{arnoldPhoton} the gluons where integrated out by (i)
including the width in the quark propagators,  
and (ii) using Feynman diagrams for
the remaining contributions. There it was shown that the leading order contributions are due to ladder diagrams with uncrossed rungs, and that ladder diagrams with
crossed rungs or vertex corrections vanish at leading order. Even
though we have not checked it explicitly, it seems to be clear that
these diagrams are part of the connected pieces.

%------------------------------------------------------------------------

\end{document}